\documentstyle[12pt]{article}
\begin{document}
\begin{center}

{\large\bf The He-McKellar-Wilkens effect for spin one particles
in non-commutative quantum mechanics} \vskip 1cm Sayipjamal
Dulat$^{a}$\footnote{dulat98@yahoo.com}, Kang
Li$^{b}$\footnote{kangli@hznu.edu.cn} and  Jianhua Wang$^{c}$
\\\vskip 1cm
 {\it\small$^a$School of Physics Science and Technology, Xinjiang
University,
Urumqi, 830046, Xinjiang, China \\
$^b $Department of Physics, Hangzhou Normal University,
Hangzhou,310036, China\\
$^c$Department of Physics, Shaanxi University  of Technology ,
Hanzhong, 723001, China}
 \vskip 0.5cm
\end{center}

\begin{abstract}
The He-McKellar-Wilkens(HMW) effect for spin one neutral particle in
non-commutative quantum mechanics is studied. By solving the Kemmer
like equations on non-commutative (NC) space and non-commutative
phase space, we obtain topological He-McKellar-Wilkens  phase on NC
space and NC phase space respectively, where the additional terms
related to the space-space and momentum-momentum non-commutativity
are given explicitly.

PACS number(s): 02.40.Gh, 11.10.Nx, 03.65.-w
\end{abstract}

\section{Introduction}

The study of physics effects on non-commutative space has attracted
much attention in recent years. Because the effects of the space
non-commutativity may become significant at the very high (Tev or
above) energy scale. Besides the field theory,  there are many
papers devoted to the study of various aspects of quantum mechanics
on NC space with usual  time coordinate\cite{NCQM}-\cite{kang}. For
example, the Aharonov-Bohm phase on NC space and  NC phase space has
been studied in Refs. \cite{AB-2}-\cite{AB-4}. The Aharonov-Casher
phase for a spin half and spin-$1$ particle on NC space and  NC
phase space has been studied in Refs.\cite{AC-NCS
spin-half}-\cite{AC-NCPS spin-one}. The equivalence of the
Aharonov-Bohm and Aharonov-Casher effects was studied in the
relativistic case for spin half particles in Ref.\cite{Hagen}. The
He-McKellar-Wilkens (HMW) effect on commutative space was firstly
discussed in 1993 by He and Meckellar\cite{HMW-spin-half-QM1} and a
year later, independently by Wilkens\cite{HMW-spin-half-QM2}. The
HMW effect corresponds to a topological phase related to a neutral
particle with non-zero electric dipole moment moving in a pure
magnetic field, and in 1998, Dowling, Willianms and Franson point
out that the HMW effect can be partially tested using metastable
hydrogen atoms\cite{HMW-spin-half-QM3}. The HMW phase for a spin
half particle on NC space and  NC phase space has been studied in
Ref.\cite{HMW-spin-half-NCQM}, but there is no discussion about the
HMW effect for spin one neutral particle in the literature, so in
this paper we study HMW effect of spin one particles.

Let's first review some basic concepts of NC quantum mechanics. In
usual commutative case the algebra of observables ${\cal A}$ is
generated by operators $x$ and $p$ satisfying the standard
commutation relations ( we take $\hbar=c=1$):
\begin{equation}\label{eq001}
[x_\nu,x_\lambda] = [p_\nu ,p_\lambda] = 0,\ [x_\nu,p_\lambda] =
i\delta_{\nu\lambda}.
\end{equation}
In NC case the commutators of the  generators of the algebra of
observables $\hat{\cal A}$  are replaced by the deformed ones, i.e.
the deformed $\hat{x}$ and $\hat{p}$ satisfy\cite{kang}:

\begin{equation}\label{eq002}
[\hat{x}_{\nu},\hat{x}_{\lambda}]=i\Theta_{\nu\lambda},~~~
[\hat{p}_{\nu},\hat{p}_{\lambda}]=i\bar{\Theta}_{\nu\lambda},~~~[\hat{x}_{\nu},\hat{p}_{\lambda}]=i
\delta_{\nu\lambda}.
\end{equation}
The NC variables can be expressed (up to some singular cases) as:
\begin{eqnarray}\label{gbshift1}
 {\hat x}_\nu = \alpha x_{\nu}-\frac{1}{2 \alpha}\Theta_{\nu\lambda}p^{\lambda},\nonumber\\
 {\hat p}_\nu = \alpha
 p_\nu+\frac{1}{2\alpha}\bar{\Theta}_{\nu\lambda}x^{\lambda}.
\end{eqnarray}
Therefore, the algebras ${\cal A}$ and $\hat{\cal A}$ are the same.
The pure states are in both cases given as wave functions $\psi(x)$
in the Hilbert space. In the NC case one has the following action
(up to terms linear in $\theta$'s): $\hat{x}\psi(x) = (\alpha
x-(1/2\alpha)\theta\wedge p)\psi(x),\ \hat{p}\psi(x) = (\alpha
p+(1/2\alpha)\bar{\theta}\wedge x)\psi(x).$ The difference between
usual quantum mechanics and NC quantum mechanics are visible only
after the choice of the Hamiltonian. If the usual hamiltonian is
given as $H =H(p,x)$, then the standard choice for the NC
Hamiltonian $\hat{H} = H(\hat{p},\hat{x})$ determines the  NC
interpretation of the NC quantum mechanics system in question. The
corresponding Schrodinger equation contains the action on a wave
function described above. Therefore there are no differences at the
level of kinematics between the usual quantum mechanics and the NC
quantum mechanics;  the difference between them is specified by
dynamics.

In NC quantum mechanics, the Schrodinger equation, as we know, can
be written as  $H(p,x)\ast \psi (p,x)=E \psi (p,x)$, from the
discussion above, this NC  Schrodinger equation can be equivalently
written by $H(\hat{p},\hat{x})\psi (p,x)=E \psi (p,x)$, i.e. the
star product can be changed into ordinary product by making shifts
\cite{kang} $x\rightarrow \hat{x}, p\rightarrow \hat{p}$. In the
case of this paper the Hamiltonian also depends on the dual of the
electromagnetic tensor $\tilde{F}$, so,  when the star product is
replaced by usual product in Schrodinger equation, the $\tilde{F}$
should be also shifted as,
\begin{equation}\label{eq5}
{\tilde F}_{\nu\lambda}\rightarrow\hat{ {\tilde {\mathcal
F}}}_{\nu\lambda}=\alpha {\tilde F}_{\nu\lambda} + \frac{1}{2\alpha}
 \Theta^{\rho\sigma} p_\rho\partial_\sigma {\tilde F}_{\nu\lambda}.
\end{equation}

 When only space-space  non-commutativity is considered we call it NC
 space, when both space-space and momentum-momentum non-commutativity
 are considered we call it NC phase space. On NC space,
 $\bar{\Theta}=0$, and it leads to $\alpha =1$,  the equations
(\ref{gbshift1}) and (\ref{eq5}) are reduced to the well know
results\cite{AB-2}:
\begin{eqnarray}
\hat{x}_\nu=x_\nu-\frac{1}{2}\Theta_{\nu\lambda}p^\lambda,
\hat{p}_\nu=p_\nu,
 \\ \label{eq4}
\tilde{F}_{\nu\lambda}\rightarrow\hat{\tilde{
F}}_{\nu\lambda}=\tilde{F}_{\nu\lambda}+\frac{1}{2}
 \Theta^{\rho\sigma} p_\rho \partial_\sigma \tilde{F}_{\nu\lambda}.
\end{eqnarray}

In this paper, first we discuss He-McKellar-Wilkens effect for a
spin-$1$ neutral particle with non-zero electric dipole moment
moving in the magnetic field  on commutative space. Then we study
the He-McKellar-Wilkens effect on non-commutative space and give a
generalized formula of HMW phase. We also give a generalized
formula of HMW phase on non-commutative phase space. Conclusion
remarks are given in the last section.

\section{The HMW effect for spin one particles in quantum mechanics }

In a similar way as in Aharonov-Bohm, Aharonov-Casher topological
effects, the He-McKellar-Wilkens effect can also be studied in
$2+1$ dimension. The ordinary configuration for HWM effect is: a
neutral particle with nonzero electric dipole moment $\mu_e$ moves
in a pure magnetic field produced by a infinitely long filament
which is uniformly charged with magnetic charge (monopoles) and
the filament is perpendicular to the plane, let us say $x-y$
plane, then the problem can be treated in $2+1$ space time. We use
the conventions $g_{\mu\nu}={\rm diag}(1,-1,-1)$.

The Dirac like equation of a spin one neutral particle with electric
dipole $\mu_e$ moving in the electromagnetic field is called Kemmer
equation and is given by\cite{kemmer-theory}
\begin{equation}\label{int-kemmer}
( i \beta^\nu \partial_\nu - {1\over 2} \mu_e S_{\lambda\rho}
\tilde{F}^{\lambda\rho} - m)\phi = 0 \;.
\end{equation}
where $\tilde{F}_{\mu\nu}=\frac{1}{2}\epsilon_{\mu\nu\alpha\beta
}F^{\alpha\beta}$ is the $3+1$ dimensional dual of the
electromagnetic field tensor. In $2+1$ dimensions, its explicit form
is
$$
\tilde F_{\mu\nu} =   \left(\matrix{ 0   &  -B^1  &  -B^2   \cr
                           B^1 &  0    &  -E^3  \cr
                           B^2 &  E^3  &  0     \cr }  \right)\;;
$$
the $10\times 10$  matrices $\beta_\nu$ are generalization of the
$4\times 4$ Dirac gamma matrices, and it can be chosen as
follows\cite{AC-QM spin1}-\cite{QFT spin1}

$$ \beta^0 =\left(\matrix{\widehat  O & \widehat O & I &  o^{\dag}\cr
                            \widehat O &\widehat O  & \widehat O & o^{\dag} \cr
                               I &\widehat O  & \widehat O & o^{\dag} \cr
                              o&o&o&0\cr} \right) \;, \ \ \ \ \ \ \
                              \\
 \beta^j=\left(\matrix{\widehat  O & \widehat  O  & \widehat  O  & -i {K}^{j \dag}\cr
                                 \widehat O & \widehat O  &S^j &  o^{\dag} \cr
                                 \widehat O &-S^j & \widehat O & o^{\dag} \cr
                                 -iK^j &o&o&0\cr}  \right)
                                 \;,
                                 $$

\noindent with $j=1,2,3$. The elements of the $10\times 10$ matrices
$\beta_\nu$ are given by the matrices

$$ \widehat O = \left(\matrix{0 & 0  & 0  \cr
                                 0& 0  & 0  \cr
                                 0 & 0 & 0 \cr}  \right) \;,\ \ \ \ \ \ \ \
                                I = \left(\matrix{1 & 0  & 0  \cr
                                                            0& 1  & 0  \cr
                                                            0 & 0 & 1 \cr}  \right) \;, $$

$$S^1 = i  \left(\matrix{0 & 0  & 0  \cr
                                 0& 0  & -1  \cr
                                 0 & 1 & 0 \cr}  \right) \;,\ \ \
                                 S^2= i  \left(\matrix{0 & 0  & 1  \cr
                                 0& 0  & 0 \cr
                                 -1 & 0 & 0 \cr}  \right)\;,\ \ \
                                 S^3= i  \left(\matrix{0 & -1  & 0  \cr
                                 1& 0  & 0 \cr
                                 0 & 0 & 0 \cr}  \right)\;,$$

$$ o = \left(\matrix{0 & 0  & 0 \cr} \right), \ \ \ K^1 = \left(\matrix{1 & 0  & 0 \cr} \right),  \ \ \
   K^2 = \left(\matrix{0 & 1  & 0 \cr} \right),  \ \ \
   K^3 = \left(\matrix{0 & 0  & 1 \cr} \right)  \;. $$

\noindent The above $\beta$ matrices satisfy the following relation
\begin{equation}\label{beta-algebra}
\beta_\nu \beta_\lambda \beta_\rho + \beta_\rho \beta_\lambda
\beta_\nu =\beta_\nu g_{\lambda\rho} + \beta_\rho g_{\nu\lambda}.
\end{equation}
And other algebraic properties of the Kemmer $\beta$-matrices were
given in Ref.\cite{kemmer-theory}.
  $S_{\lambda\rho}$ is the Dirac
$\sigma_{\lambda\rho}$ like spin operator, which can be defined as
\begin{equation}\label{spin-s}
S_{\lambda\rho} = \frac{1}{2}(\beta_\lambda\beta_\rho
-\beta_\rho\beta_\lambda)\;.
\end{equation}

The solution of the Kemmer equation can be written in the following
form
\begin{equation}\label{solution-free-kemmer}
\phi = e^{-i\xi_3 \int^r  {\mathbf a} \cdot \;d\mathbf r}\;\phi_0,
\end{equation}
where $\phi_0$ is a solution of the free Kemmer equation; the spin
one pseudo-vector operator $\xi_\nu$ in (\ref
{solution-free-kemmer}) is defined as
\begin{equation}\label{spin xi}
\xi_\nu
=\frac{i}{2}\varepsilon_{\nu\lambda\rho\sigma}\beta^\lambda\beta^\rho\beta^\sigma\;,
\end{equation}
where $\varepsilon_{\nu\lambda\rho\sigma}$ is the Levi-Civita
symbol in four dimensions. Now we need to find the explicit form
of the vector ${\mathbf a}$ in (\ref{solution-free-kemmer}). To do
this, first we write the free Kemmer equation for $\phi_0$ in
terms of $\phi$
\begin{equation}\label{free-kemmer-2}
( i \beta^\nu \partial_\nu- m)\;e^{i\xi_3 \int^r {\mathbf a} \cdot
\;d{\mathbf r}}\;\phi= 0
\end{equation}
We impose the following two conditions in order to have the
equivalence of (\ref{int-kemmer}) and (\ref{free-kemmer-2})
\begin{equation}\label{condition-1}
e^{-i\xi_3 \int^r {\mathbf a} \cdot \;d{\mathbf
r}}\;\beta^\nu\;e^{i\xi_3 \int^r {\mathbf a} \cdot \;d{\mathbf
r}}= \beta^\nu,
\end{equation}
and
\begin{equation}\label{condition-2}
\beta^\nu\xi_3 a_\nu \phi ={1\over 2} \mu_e S_{\lambda\rho}
\tilde{F}^{\lambda\rho}\phi = \mu_e S_{0l} \tilde{F}^{0l}\phi.
\end{equation}
By  comparing (\ref{condition-1}) with the Baker-Housdorf formula
\begin{equation}\label{housdorf}
e^{-i\lambda\xi_3}\;\beta^\nu \;e^{i\lambda\xi_3} = \beta^\nu +
\wp(-i\lambda)[\xi_3,\beta^\nu] +
\frac{1}{2!}\wp(-i\lambda)^2[\xi_3,[\xi_3,\beta^\nu]]\ldots,
\end{equation}
we get, $ [\xi_3,\beta^\nu]=0$, where $\wp$ stands for  path
ordering of the integral in the phase. If $\nu \neq 3$ this
commutation relation is automatically satisfied, however, for
$\nu=3$, by using (\ref{beta-algebra}) and (\ref{spin xi}), we
find that the commutator does not vanish. Thus  to satisfy the
first condition we restrict the particle in $x-y$ plane, that is,
$B_z=0$. In particular $\partial_3\phi=0$ and $a_3=0$. From
(\ref{condition-2}), by using (\ref{beta-algebra}), (\ref{spin-s})
and (\ref{spin xi}), one obtains
\begin{equation}
 a_l = 2\mu_e \varepsilon_{lk}B_k \;,\hspace{2cm}  l,k =
1,2\;.
\end{equation}
Thus the HMW phase for a neutral spin one particle moving in a
$2+1$ space time under the influence of a pure magnetic field
produced by an infinitely long filament which is uniformly charged
with magnetic monopoles is given by
\begin{equation}\label{HMW-phase}
\phi_{HMW} = \xi_3\oint {\mathbf a} \cdot\; d{\mathbf r}= 2\mu_e
\xi_3\varepsilon^{lk}\oint B_ldx_k= 2\mu_e \xi_3\oint (-B_1 dx_2 +
B_2 dx_1)\;.
\end{equation}
 The above equation can also be written as
\begin{equation}\label{HMW-phase-1}
\phi_{HMW} = \xi_3\oint {\mathbf a} \cdot\; d{\mathbf r}=
\xi_3\int_S({\mathbf\nabla}\times {\mathbf a})\cdot d{\mathbf S} =
2\mu_e\xi_3 \int_S ({\mathbf \nabla}\cdot{\mathbf B}) dS =
2\mu_e\xi_3 \lambda_m\;,
\end{equation}
where $\lambda_m$ is the magnetic charge density of the filament.
This spin one HMW phase is also purely quantum mechanical effect
and has no classical interpretation.   One may note that the HMW
phase for spin one particles is exactly the same as those for spin
half, except that the spin operator and spinor have changed. The
factor of two shows that the phase is twice that accumulated by a
spin half particle with the same electric dipole moment, in the
same magnetic field.

\section{ HMW effect for spin one particles in non-commutative  quantum mechanics}

 In this section we study HMW effect for spin one particles both on NC space and NC phase space.
 By replacing the usual product in (\ref{int-kemmer}) with a
star product (Moyal-Weyl product), the Kemmer equation for a spin
one neutral particle with a electric dipole moment $\mu_e$, on NC
space, can be written as

\begin{equation}\label{int-kemmer-NC-1}
( i \beta^\nu \partial_\nu - {1\over 2} \mu_e S_{\lambda\rho}
\tilde {F}^{\lambda\rho} - m)\ast\phi = 0,
\end{equation}
By (\ref{eq4}), we replace the star product in
(\ref{int-kemmer-NC-1}) with ordinary product, then the Kemmer
equation on NC space has the form
\begin{equation}\label{int-kemmer-NC}
( i \beta^\nu \partial_\nu - {1\over 2} \mu_e S_{\lambda\rho}
\hat{\tilde{F}}^{\lambda\rho} - m)\phi = 0.
\end{equation}
In a similar way as the commuting space, the solution of the above
equation can also be written as
\begin{equation}\label{phi-NC}
\phi = e^{-i\xi_3 \int^r \hat{\mathbf a} \cdot \;d\mathbf
r}\;\phi_0 \;.
\end{equation}
To determine $\hat {\mathbf a}$ we write the free Kemmer equation
as
\begin{equation}\label{free-kemmer-NC-2}
( i \beta^\nu \partial_\nu- m)\;e^{i\xi_3 \int^r \hat{\mathbf a}
\cdot \;d\mathbf r}\;\phi= 0 \;.
\end{equation}
The equivalence  of (\ref{int-kemmer-NC}) and
(\ref{free-kemmer-NC-2}) gives the following two conditions
\begin{equation}\label{condition-1-NC}
e^{-i\xi_3 \int^r \hat{\mathbf a} \cdot \;d\mathbf
r}\;\beta^\nu\;e^{i\xi_3 \int^r \hat{\mathbf a}\cdot \;d\mathbf
r}= \beta^\nu
\end{equation}
and
\begin{equation}\label{condition-2-NC}
\beta^\nu\xi_3 \hat a_\nu \phi ={1\over 2} \mu_e S^{\lambda\rho}
\hat{ \tilde{F}}_{\lambda\rho}\phi = \mu_e S^{0l}
\hat{\tilde{F}}_{0l}\phi \;.
\end{equation}
By using  (\ref{housdorf}), the first condition
(\ref{condition-1-NC}) implies that, $ [\xi_3,\beta^\nu]=0$. If
$\nu \neq 3$ then this commutation relation is automatically
satisfied, however, for $\nu=3$, by using (\ref{beta-algebra}) and
(\ref{spin xi}), one finds that the commutator does not vanish.
Therefore in order to fulfil the first condition we restrict the
particle in $2+1$ space-time. In particular $\partial_3\phi=0$ and
$\hat a_3=0$. From  (\ref{condition-2-NC}), and by using
(\ref{beta-algebra}), (\ref{spin-s}) and (\ref{spin xi}), we
obtain
\begin{eqnarray}
\hat a_1 &=& 2\mu_e \hat {\tilde F}_{02}= 2\mu_e {\tilde F}_{02} +
2\mu_e \frac{1}{2}\Theta^{ij}p_i\partial_j {\tilde F}_{02} =2\mu_e
B_2 + \mu_e
\theta \varepsilon^{ij} p_i\partial_j B_2\nonumber\\
\hat a_2 &=& -2\mu_e \hat {\tilde F}_{01}= -2\mu_e {\tilde F}_{01}
- 2\mu_e \frac{1}{2}\Theta^{ij}p_i\partial_j {\tilde
F}_{01}=-2\mu_e B_1 + \mu_e \theta \varepsilon^{ij} p_i\partial_j
B_1
\end{eqnarray}
with $\Theta^{ij}=\theta\epsilon^{ij}$, $\Theta^{0\mu}=\Theta^{\mu
0}=0$; $\epsilon^{ij}=-\epsilon^{ji}$, $\epsilon^{12}=+1$. Thus
the HMW phase for a neutral spin one particle moving in a $2+1$
non-commutative space under the influence of a pure magnetic field
produced by an infinitely long filament which is uniformly charged
with magnetic monopoles,  is
\begin{equation}\label{HMW-phase-NC}
\hat\phi_{HMW} = \xi_3\oint \hat{\mathbf a} \cdot\; d{\mathbf
r}=2\mu_e \xi_3\varepsilon^{lk}\oint B_ldx_k + \mu_e \xi_3\theta
\varepsilon^{ij}\varepsilon^{lk}\oint p_i\partial_j B_l dx_k\;.
\end{equation}
In a similar way as in  spin half case \cite{HMW-spin-half-NCQM},
the momentum on NC space for a spin-$1$ neutral particle can also
be written as
\begin{equation}\label{eq29}
p_i=mv_i+(\vec B\times\vec \mu)_i+\mathcal{O}(\theta),
\end{equation}
where $\vec{\mu}=2\mu_e \vec S$, and $\vec S$ is the spin operator
of the spin one. By inserting (\ref{eq29}) into
(\ref{HMW-phase-NC}), we have
\begin{equation}\label{eq31}
\hat{\phi}_{HMW}=\phi_{HMW}+\delta\phi_{NCS},
\end{equation}
where $\phi_{HMW}$ is the HMW phase (\ref{HMW-phase}) on commuting
space; the  additional phase $\delta \phi_{NCS}$, related to the
non-commutativity of space, is given by
\begin{eqnarray}\label{eq32}
\delta\phi_{NCS}=\mu_e \xi_3\theta
\varepsilon^{ij}\varepsilon^{lk}\oint [k_i - (\vec\mu \times \vec
B)_i]\partial_j B_l dx_k
\end{eqnarray}
where $k_i=mv_i$ is the wave number; the $\xi_3$ in the phase
represents the spin degrees of freedom. If the spin of the neutral
particle along the $z$ direction, namely, $\vec{\mu}=2 \mu_e s_3
\hat{\vec{k}}$, then the above equation takes the form
\begin{eqnarray}
\delta\phi_{NCS}=\mu_e \xi_3\theta
\varepsilon^{ij}\varepsilon^{lk}\oint [k_i - 2\mu_e
s_3(\hat{\vec{k}} \times \vec B)_i]\partial_j B_l dx_k \;,
\end{eqnarray}
where $\hat{\vec{k}}$ is a unite vector in the z direction; $s_3 =
1, 0, -1$.

Now we discuss the HMW phase on NC phase space. From
(\ref{gbshift1}), (\ref{eq5}) and (\ref{int-kemmer-NC-1}), the
Kemmer equation for HMW problem on NC phase space has the form
\begin{equation}\label{int-kemmer-NCPS-2}
( - \beta^\nu \hat{p}_\nu
 -
{1\over 2} \mu_e S_{\lambda\rho} \hat {\tilde {\mathcal
F}}^{\lambda\rho} - m )\phi = 0.
\end{equation}
Because $\alpha\neq 0$, the above  equation can be written as
\begin{equation}\label{int-kemmer-NCPS-3}
\Big( - \beta^\nu p_\nu -
\frac{1}{2\alpha^2}\beta^\nu\bar{\Theta}_{\nu\lambda}x^{\lambda} -
{1\over 2} \mu_e S_{\lambda\rho} ( {\tilde F}^{\lambda\rho} +
\frac{1}{2\alpha^2}
 \Theta^{\sigma\tau} p_\sigma\partial_\tau {\tilde F}^{\lambda\rho} ) -
m^\prime \Big)\phi = 0.
\end{equation}
where $m'=m/\alpha$. We write the above equation  in the following
form
\begin{equation}\label{free-kemmer-NCPS}
( - \beta^\nu p_\nu  - m^\prime)\;e^{\frac{i}{2\alpha^2 } \int^r
\bar{\Theta}_{\nu\lambda} x^{\lambda}dx^\nu  + i\xi_3 \int^r
\hat{\mathbf a}^\prime \cdot \;d\mathbf{r}}\;\phi= 0 \;.
\end{equation}
To have the equivalence  of (\ref{int-kemmer-NCPS-3}) and
(\ref{free-kemmer-NCPS}), we impose the following two conditions

\begin{equation}\label{condition-1-NCPS}
e^{-i\xi_3 \int^r  \hat{\mathbf a }^\prime \cdot \;d{\mathbf
r}}\;\beta^\nu\;e^{i\xi_3 \int^r \hat{\mathbf a}^\prime\cdot
\;d{\mathbf r}}= \beta^\nu \;,
\end{equation}
and
\begin{equation}\label{condition-2-NCPS}
-\beta^\nu \xi_3 \hat{ a}^\prime_\nu \phi ={1\over 2\alpha} \mu_e
S_{\lambda\rho} \hat {\tilde F}^{\lambda\rho}\phi =
\frac{\mu_e}{\alpha} S_{0l} \hat{\tilde F}^{0l}\phi \;.
\end{equation}
In an analogous way as in NC space, from (\ref{condition-1-NCPS})
and (\ref{condition-2-NCPS}) one obtains
\begin{eqnarray}
\hat {a }^\prime_1 &=& \frac{2\mu_e}{\alpha} \hat {\tilde F}^{02}=
2\mu_e {\tilde F}^{02} + 2\mu_e
\frac{1}{2\alpha^2}\Theta^{ij}p_i\partial_j {\tilde F}^{02}
=2\mu_e B_2 + \frac{\mu_e
\theta }{\alpha^2}\varepsilon^{ij} p_i\partial_j B_2\;,\nonumber\\
\hat {a}^\prime_2 &=& -\frac{2\mu_e}{\alpha} \hat {\tilde F}^{01}=
-2\mu_e {\tilde F}^{01} - 2\mu_e
\frac{1}{2\alpha^2}\Theta^{ij}p_i\partial_j {\tilde F}^{01}=
-2\mu_e B_1 - \frac{\mu_e \theta }{\alpha^2} \theta
\varepsilon^{ij} p_i\partial_j B_1\;,\nonumber\\
\hat {a}^\prime_3 &=& 0\;.
\end{eqnarray}
Thus the HMW phase for a neutral spin one particle moving in a
$2+1$ non-commutative phase space  under the influence of a pure
magnetic field produced by a  infinitely long filament, which is
uniformly charged with  magnetic monopoles, and which is
perpendicular to the plane, is given by
\begin{eqnarray}\label{eq24}
\hat\phi_{HMW} &=& \frac{1}{2\alpha^2 } \oint
\bar{\Theta}_{\nu\lambda} x^{\lambda}dx^\nu  + \xi_3\oint
\hat{{\mathbf a}}^\prime \cdot\; d{\mathbf r}\nonumber\\& =&
\frac{\theta}{2\alpha^2 } \oint \varepsilon^{ij} x_j dx_i + 2\mu_e
\xi_3\varepsilon^{lk}\oint B_ldx_k + \mu_e \xi_3\frac{\theta
}{\alpha^2}\varepsilon^{ij}\varepsilon^{lk}\oint p_i\partial_j B_l
dx_k
\end{eqnarray}
By using $p_i=k^\prime_i+(\vec B\times\vec
\mu)_i+\mathcal{O}(\theta)$, and $k^\prime_i=m^\prime_i v_i$,
$\vec{\mu}=2\mu_e \vec S$, one obatains
\begin{equation}\label{HMW-NCPS}
\hat{\varphi}_{HMW}=\phi_{HMW}+\delta\phi_{NCS}+\delta\phi_{NCPS},
\end{equation}
where $\phi_{HMW}$  is the HMW phase (\ref{HMW-phase}) on commuting
space; $\delta\phi_{NCS}$ is the space-space non-commuting
contribution to the HMW phase (\ref{HMW-phase}) , and its explicit
form is given in (\ref{eq32}); the last term $\delta\phi_{NCPS}$ is
the momentum-momentum non-commuting contribution to the HMW phase,
and it has the form
\begin{eqnarray}\label{HMW-NCPS-correction}
\delta\phi_{NCPS} &=& \frac{\bar{\theta}}{2\alpha^2 } \oint
\varepsilon^{ij} x_j dx_i +(\frac{1}{\alpha^2} -1)\mu_e
\xi_3\theta \varepsilon^{ij}\varepsilon^{lk}\oint k^\prime_i
\partial_j B_l dx_k \nonumber\\  & - & (\frac{1}{\alpha^2} -1)\mu_e
\xi_3\theta \varepsilon^{ij}\varepsilon^{lk}\oint(\vec\mu \times
\vec B)_i\partial_j B_l dx_k \;,
\end{eqnarray}
which represents the non-commutativity of the momenta. The first
term in (\ref{HMW-NCPS-correction}) comes from the
momentum-momentum non-commutativity; the second term is a velocity
dependent correction and does not have the topological properties
of the commutative HMW effect and could modify the phase shift;
the third term is a correction to the vortex and does not
contribute to the line spectrum. In two dimensional
non-commutative plane,
$\bar{\Theta}_{ij}=\bar{\theta}\epsilon_{ij}$, and the two NC
parameters $\theta$ and $\bar{\theta}$ are related by
$\bar{\theta}=4\alpha^2 (1-\alpha^2)/\theta$ \cite{kang}. When
$\alpha=1$, which will lead to $\bar{\theta}_{ij}=0$, then the HMW
phase on NC phase space will returns to the HMW phase on NC space,
i.e. $\delta\phi_{NCPS}=0$ and Eq.(\ref{HMW-NCPS}) will changes
into Eq.(\ref{eq31}).

\section{Conclusion remarks }
There are two methods, namely, star product and shift method, to
study physical effects on NC space and NC phase space. In this
paper, first study the HMW effect in quantum mechanics. Then by
using the shift method we give the NC space corrections to
 the topological phase of the HMW effect for a
spin one neutral particle. Furthermore, by considering the
momentum-momentum non-commutativity we obtain the NC phase space
corrections to
 the topological phase of the HMW effect for a
spin one neutral particle. We note that the corrections
(\ref{eq32}) and (\ref{HMW-NCPS-correction}) to the topological
phase (\ref{HMW-phase}) or (\ref{HMW-phase-1}) of the HMW effect
for a spin one neutral particle both on NC space  and NC phase
space can be obtained  from spin half corrections
\cite{HMW-spin-half-NCQM} through the replacement
$\frac{1}{2}\gamma^0\sigma^{12}\longrightarrow \xi_3$. One may
conclude that, apart from the spin operators, the NC HMW phase for
a higher spin  neutral particle is the same as those for spin half
and spin one case in non-commutative quantum mechanics.

The method we use in this paper may also be employed to other
physics problem on NC space and NC phase space.

\section{Acknowledgments}
 This work is supported  by the
National Natural Science Foundation of China (10665001, and 10575026
as well as 10447005).

\end{document}